\documentclass[12pt,preprint]{aastex}

\shorttitle{Ionized Warped Disk and Disk Wind in MonR2-IRS2}
\shortauthors{Jim\'enez-Serra et al.}

\begin{document}


\title{The ionized warped disk and disk wind of the massive protostar Monoceros R2-IRS2 seen with ALMA}


\author{Izaskun Jim\'{e}nez-Serra\altaffilmark{1}, Alejandro B\'aez-Rubio\altaffilmark{1}, Jes\'us Mart\'{\i}n-Pintado\altaffilmark{1}, Qizhou Zhang\altaffilmark{2}, and V\'{\i}ctor M. Rivilla\altaffilmark{3}}

\altaffiltext{1}{Centro de Astrobiolog\'{\i}a (CSIC-INTA), Ctra. de Torrej\'on a Ajalvir km 4, 28850, Torrej\'on de Ardoz, Spain; ijimenez@cab.inta-csic.es}
\altaffiltext{2}{Center for Astrophysics | Harvard \& Smithsonian, 60 Garden St., Cambridge, MA 02138 USA}
\altaffiltext{3}{INAF-Osservatorio Astrofisico di Arcetri, Largo E. Fermi 5, I-50125 Firenze, Italy}

\begin{abstract}

Theories of massive star formation predict that massive protostars accrete gas through circumstellar disks. Although several cases have been found already thanks to high-angular resolution interferometry, it remains unknown the internal physical structure of these disks and, in particular, whether they present warps or internal holes as observed in low-mass proto-planetary disks. Here, we report very high angular resolution observations of the H21$\alpha$ radio recombination line carried out in Band 9 with the Atacama Large Millimeter/submillimeter Array (beam of 80$\,$mas$\times$60$\,$mas, or 70$\,$au$\times$50$\,$au) toward the IRS2 massive young stellar object in the Monoceros R2 star-forming cluster. The H21$\alpha$ line shows maser amplification, which allows us to study the kinematics and physical structure of the ionised gas around the massive protostar down to spatial scales of $\sim$1-2$\,$au. Our ALMA images and 3D radiative transfer modelling reveal that the ionized gas around IRS2 is distributed in a Keplerian circumstellar disk and an expanding wind. The H21$\alpha$ emission centroids at velocities between $-$10 and 20$\,$km$\,$s$^{-1}$ deviate from the disk plane, suggesting a warping for the disk. This could be explained by the presence of a secondary object (a stellar companion or a massive planet) within the system. The ionized wind seems to be launched from the disk surface at distances $\sim$11$\,$au from the central star, consistent with magnetically-regulated disk wind models. This suggests a similar wind launching mechanism to that recently found for evolved massive stars such as MWC349A and MWC922.  

\end{abstract}


\section{Introduction}

The formation processes of massive stars (with masses $\geq$8$\,$M$_\odot$) are under debate. Massive stars could form either by the monolithic collapse of a turbulent molecular core \citep{mckee02}, by the competitive accretion of core materials \citep{bonnell06}, or by Bondi-Hoyle accretion onto the central star \citep{keto07}. In all these theories, gas accretion occurs through a circumstellar disk coupled with an expanding wind/outflow (which removes angular momentum), in a scaled-up version of low-mass star formation. High-angular resolution interferometry has unveiled several cases of circumstellar disks around massive protostars such as Cepheus A HW2 \citep[][]{patel05,jimenez-serra07,jimenez-serra09}, IRAS 20126+4104 \citep[][]{cesaroni14,chen16}, G17.64+0.16 \citep[][]{maud19}, or GGD 27-MM1 \citep[][]{anez20}. However, the level of detail in these observations only goes down to 40$\,$au \citep[][]{maud19,anez20}, insufficient to probe the innermost structure of these disks. Hints of holes have recently been reported \citep{maud19}, but it remains unknown whether warps exist within these disks as found for their low-mass counterparts \citep[][]{sakai19,sai20,kraus20_1}. The launching process of winds/outflows has not been witnessed either during the protostellar phase of massive stars\footnote{The protostellar nature of the emission line star MWC349A is highly debated \citep[][]{hartmann80,strelnitski13,gvaramadze12,zhang17}. The latest results suggest that this object is an evolved B[e] supergiant \citep[][]{kraus20_2}.}.  

Hydrogen radio recombination lines (RRLs) probe the kinematics of the ionized gas in ultracompact (UC) HII regions during the process of massive star formation \citep[][]{churchwell02}. About 30\% of these regions show broad RRLs with linewidths $\geq$60-80$\,$km$\,$s$^{-1}$, which exceed those produced by just thermal broadening \citep[FWHM$\sim$30$\,$km$\,$s$^{-1}$;][]{churchwell02}. These regions show elongated morphologies and thus, broad RRL emission likely arise from ionized winds \citep[][]{jaffe99}. 

Non-LTE effects such as maser amplification are expected in RRLs. These effects are pronounced in transitions at sub-millimeter wavelengths when electron densities reach values $n_e$$\sim$10$^6$-10$^8$$\,$cm$^{-3}$ \citep[Figure 5 in][]{strelnitski96}. However, RRL masers are rare. Only a few objects have been detected so far such as MWC349A \citep[][]{martin-pintado89a,baez13}, eta Carinae \citep[][]{cox95}, or the evolved B[e]-type massive star MWC922 \citep[][]{sanchez-contreras19}. Thanks to their brightness, the kinematics of the innermost ionized regions toward these evolved objects can be measured with accuracies down to a few au \citep[][]{weintroub08,martin-pintado11,baez13,zhang17}. This has revealed that they not only present circumstellar disks rotating following a Keplerian law, but also that ionized winds are launched from their disks at radii $\lesssim$25-30$\,$au \citep[][]{martin-pintado11,sanchez-contreras19}. Unfortunately, similar studies do not exist for massive protostellar objects. 

Monoceros R2 (hereafter MonR2) is a massive star-forming cluster located at a distance of 893$\,$pc \citep[][]{dzib16}. It hosts a blister-type HII region and a cluster of IR sources \citep[][]{massi85,wood89,carpenter97,trevino-morales19}. Among them, IRS2 is a compact and massive young stellar object (YSO) with a luminosity $\sim$0.5-1$\times$10$^4$$\,$L$_\odot$ \citep{howard94,alvarez04}. High-angular resolution observations carried out with the Submillimeter Array (SMA) revealed that MonR2-IRS2 is a UC HII region with its RRLs at $\lambda$$\leq$0.85$\,$mm experiencing maser amplification. 
However, these observations were unable to resolve the internal structure of the source \citep[][]{jimenez-serra13}.

We present high-angular resolution observations (80$\,$mas$\times$60$\,$mas, 70$\,$au$\times$50$\,$au) of the H21$\alpha$ RRL carried out with the Atacama Large Millimeter/submillimeter Array (ALMA) toward the MonR2-IRS2 UC HII region. Our H21$\alpha$ images show that the ionized gas in MonR2-IRS2 is distributed in a warped Keplerian disk and an ionized wind that is launched at radii $\sim$11$\,$au.

\section{Observations}
\label{obs}

The H21$\alpha$ RRL at 662.40416$\,$GHz was observed on 25 September 2015 with ALMA in Band 9 (project 2012.1.00522.S) using the 12m array (baselines from 43$\,$m to 2.27$\,$km). 
Observations were performed in dual polarization mode using a spectral bandwidth of 1.875$\,$GHz and a spectral resolution of 976$\,$kHz (0.44$\,$km$\,$s$^{-1}$). As calibrators, J0522-3627 was used for bandpass calibration, J0423-013 for flux calibration, and J0607-0834 for phase and amplitude calibration. 

The calibrated dataset was obtained by running the original pipeline reduction scripts, and the data were additionally self-calibrated using the CASA (the Common Astronomy Software Applications) package version 5.0.0. The MonR2-IRS2 spectra are clean from molecular emission and thus, the 0.4$\,$mm continuum map was generated by using the CASA task {\it uvcontsub} across 5.61$\,$GHz of line-free bandwidth. The angular resolution of the 0.4$\,$mm continuum image was 60$\,$mas$\times$50$\,$mas, P.A.=78$^{\circ}$ and the rms noise level was 1.9$\,$mJy$\,$beam$^{-1}$. The H21$\alpha$ image was obtained from the continuum-subtracted datacube after smoothing the velocity resolution to 1$\,$km$\,$s$^{-1}$. The beam and rms noise level of the H21$\alpha$ image were 80$\,$mas$\times$60$\,$mas, P.A.=$-$88$^{\circ}$, and 30$\,$mJy$\,$beam$^{-1}$ in 1$\,$km$\,$s$^{-1}$-channels, respectively. 

\section{Results}
\label{results}

\subsection{0.4$\,$mm continuum emission}
\label{continuum}

\begin{figure}
\begin{center}
\includegraphics[angle=270,width=1.0\textwidth]{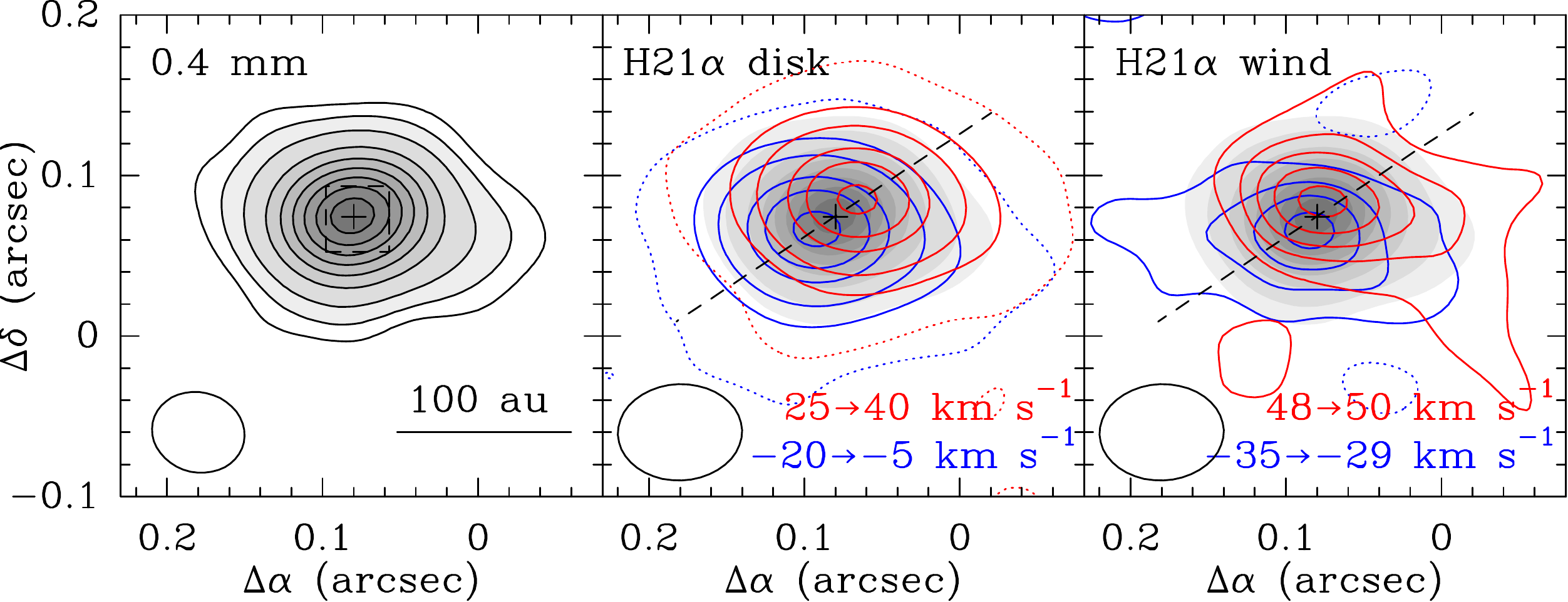}
\caption{Left panel: 0.4$\,$mm continuum map measured toward MonR2-IRS2 with ALMA. Contours correspond to 3$\sigma$, 6$\sigma$, 12$\sigma$, 24$\sigma$, 36$\sigma$, 48$\sigma$, 60$\sigma$, 72$\sigma$, and 84$\sigma$, with 1$\sigma$=1.9$\,$mJy$\,$beam$^{-1}$. Cross indicates the continuum peak. Dashed box shows the region of the H21$\alpha$ centroid map of Figure$\,$\ref{centroid_map_figure}. Middle panel: H21$\alpha$ integrated intensity maps obtained for the ionized disk between 25 and 40$\,$km$\,$s$^{-1}$ (red) and $-$20 and $-$5$\,$km$\,$s$^{-1}$ (blue). Contours correspond to the 3$\sigma$ (dotted lines), 20$\sigma$, 40$\sigma$, 80$\sigma$, 110$\sigma$, and 150$\sigma$ levels, with 1$\sigma$=160$\,$mJy$\,$beam$^{-1}$$\,$km$\,$s$^{-1}$, for both the red- and blue-shifted parts of the disk. Grey scale shows the continuum emission. Dashed line indicates the disk plane. Right panel: H21$\alpha$ maps of the high-velocity gas obtained between 48 and 50$\,$km$\,$s$^{-1}$ (contours at 2$\sigma$, 5$\sigma$, 7.5$\sigma$, and 10$\sigma$, with 1$\sigma$=80$\,$mJy$\,$beam$^{-1}$$\,$km$\,$s$^{-1}$), and between $-$35 and $-$29$\,$km$\,$s$^{-1}$ (contours at 2$\sigma$, 5$\sigma$, 7.5$\sigma$, and 10$\sigma$, and 12.5$\sigma$, with 1$\sigma$=42$\,$mJy$\,$beam$^{-1}$$\,$km$\,$s$^{-1}$). Grey scale and dashed line are as in the middle panel. ALMA beams are shown in the lower left corner of each panel.}
\label{fig1}
\end{center}
\end{figure}

Figure$\,$\ref{fig1} (left panel) reports the 0.4$\,$mm continuum map obtained toward MonR2-IRS2 (black contours and grey scale). The source is barely resolved at the resolution of ALMA (deconvolved size 62$\,$mas$\times$43$\,$mas, or 55$\,$au$\times$38$\,$au). Its morphology clearly deviates from gaussianity at the low-emission level ($\leq$12$\sigma$, with 1$\sigma$=1.9$\,$mJy$\,$beam$^{-1}$), likely due to the presence of a wind (Section$\,$\ref{model}). By fitting a 2D gaussian to the continuum emission, the peak coordinates of MonR2-IRS2 are $\alpha$(J2000)=06$^h$07$^m$45.8034$^s$ ($\pm$0.0008$^s$), $\delta$(J2000)=$-$06$^\circ$22$'$53.5155$"$ ($\pm$0.0005$"$), which are consistent with those previously measured \citep[][]{jimenez-serra13}. 

The integrated continuum flux at 0.4$\,$mm is (345$\pm$10)$\,$mJy. This value is larger than that expected from the spectral index of $\alpha$=$-$0.16$\pm$0.2 measured by the SMA \citep[with $S\propto\nu^\alpha$;][]{jimenez-serra13}\footnote{The uncertainty in $\alpha$ considers the rms noise and systematic errors ($\leq$5\%) in the SMA data. Note that the MonR2-IRS2 continuum flux in both SMA VEX and COM images differs by 2\% \citep[Table$\,$2 in][]{jimenez-serra13}.}. The free-free contribution is 130$\pm$30$\,$mJy and thus, while the continuum emission at $\lambda$$\geq$0.85$\,$mm is entirely due to ionized gas, the $\lambda$=0.4$\,$mm continuum flux is dominated by dust. The mass for the neutral disk can then be estimated using a dust flux of 215$\pm$32$\,$mJy and by assuming dust temperatures T$_{dust}$=500-1200$\,$K \citep{howard94}, a gas-to-dust mass ratio R=100, and a dust opacity $\kappa_{dust}$=25.8$\,$cm$^2$$\,$g$^{-1}$ \citep[for H$_2$ gas densities of 10$^8$$\,$cm$^{-3}$ and no ices;][]{ossenkopf94}. The neutral disk mass is $\sim$2-4$\times$10$^{-4}$$\,$M$_\odot$ with a 15\% uncertainty.  

\subsection{H21$\alpha$ maser emission}
\label{h21a}

\begin{figure}
\begin{center}
\includegraphics[angle=270,width=1.0\textwidth]{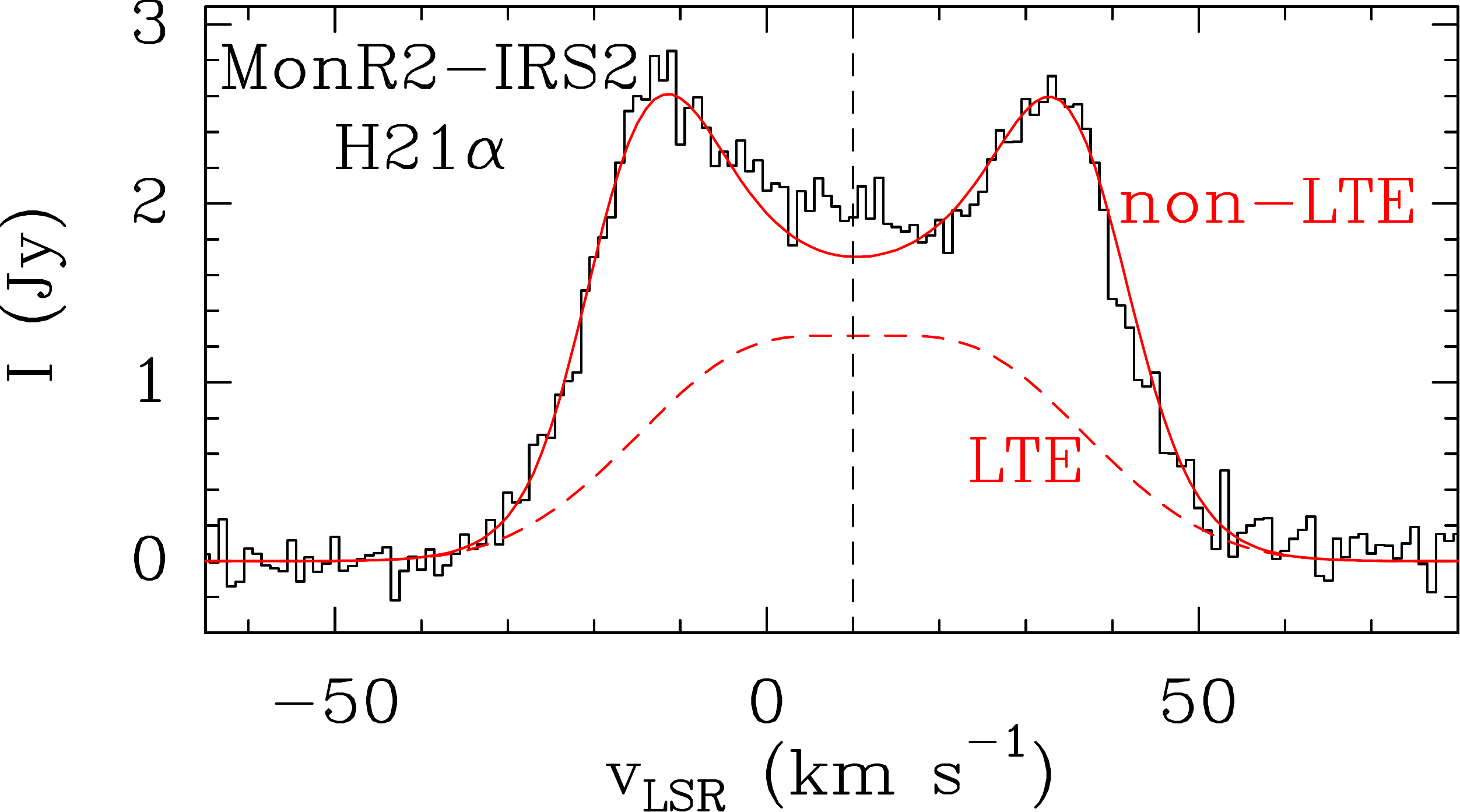}
\caption{H21$\alpha$ RRL profile (histogram) integrated over the MonR2-IRS2 source. The spectra modelled using the MORELI code for LTE (red dashed line) and non-LTE conditions (red solid line), are also shown (Section$\,$\ref{model}). Vertical dashed line indicates the systemic velocity at V$_{LSR}$=10$\,$km$\,$s$^{-1}$.}
\label{spectrum}
\end{center}
\end{figure}

In Figure$\,$\ref{spectrum}, we present the H21$\alpha$ spectrum integrated over the MonR2-IRS2 continuum source (Figure$\,$\ref{fig1}). Being consistent with the H30$\alpha$ and H26$\alpha$ lines measured with the SMA \citep[][]{jimenez-serra13}, the H21$\alpha$ transition shows two emission peaks shifted by $\pm$22$\,$km$\,$s$^{-1}$ with respect to MonR2-IRS2's radial velocity at 10$\,$km$\,$s$^{-1}$ \citep[][]{torrelles83}. The intensity of the two peaks is about the same ($\sim$2.8$\,$Jy) and a factor of 10 higher than that of the H30$\alpha$ RRL \citep[$\sim$0.28$\,$Jy$\,$beam$^{-1}$; Table$\,$2 in][]{jimenez-serra13}\footnote{Note that the SMA beams ranged between 0.4$"$-2.4$"$ and therefore, the SMA data include all emission from the MonR2-IRS2 source (size $\leq$60$\,$mas; Section$\,$\ref{continuum}).}. This intensity increase is expected since the line-to-continuum flux ratio (LTCR)\footnote{The LTCR is defined as $\frac{\Delta v\,T_L}{T_C}$, with $\Delta v$ and $T_L$ being respectively the linewidth and peak intensity of the RRL, and $T_C$ the free-free continuum flux.} of RRLs under LTE conditions and optically thin emission, is known to increase with frequency as $\nu$$^{1.1}$. The LTCR is predicted to be $\sim$206$\,$km$\,$s$^{-1}$ for the H21$\alpha$ line. However, the measured one
is $\sim$1130$\,$km$\,$s$^{-1}$, which clearly exceeds the LTCR predicted under LTE.  This, together with our RRL radiative transfer calculations of Section$\,$\ref{model}, confirm that the RRLs at $\lambda$$\leq$0.85$\,$mm toward MonR2-IRS2 are masers.  

Figure$\,$\ref{fig1} reports the integrated intensity maps for the H21$\alpha$ blue- and red-shifted peaks from -20 to -5$\,$km$\,$s$^{-1}$ and from 25 to 40$\,$km$\,$s$^{-1}$, respectively (middle panel), and for the high-velocity blue and red wings (from -35 to -29$\,$km$\,$s$^{-1}$ and from 48 to 50$\,$km$\,$s$^{-1}$ respectively; right panel). The H21$\alpha$ maps of the emission peaks follow the morphology of the continuum (left panel) but their maxima are displaced by $\sim$0.02$"$ ($\sim$18$\,$au) from the continuum peak. While the red-shifted emission peaks toward the north-west of MonR2-IRS2, the blue-shifted part peaks toward the south-east. As shown in Section$\,$\ref{model}, this emission is well reproduced by a Keplerian ionized disk. For the high-velocity wings, the H21$\alpha$ line follows a similar kinematic trend (red-shifted gas in the north-west, blue-shifted emission in the south-east), but its morphology appears slightly elongated in the direction perpendicular to the disk (see dashed line in Figure$\,$\ref{fig1}, right panel), as if an ionized wind were launched from the disk (Sections$\,$\ref{centroid} and \ref{model}). 

\subsection{H21$\alpha$ centroid map}
\label{centroid}

\begin{figure}
\begin{center}
\includegraphics[angle=270,width=1.0\textwidth]{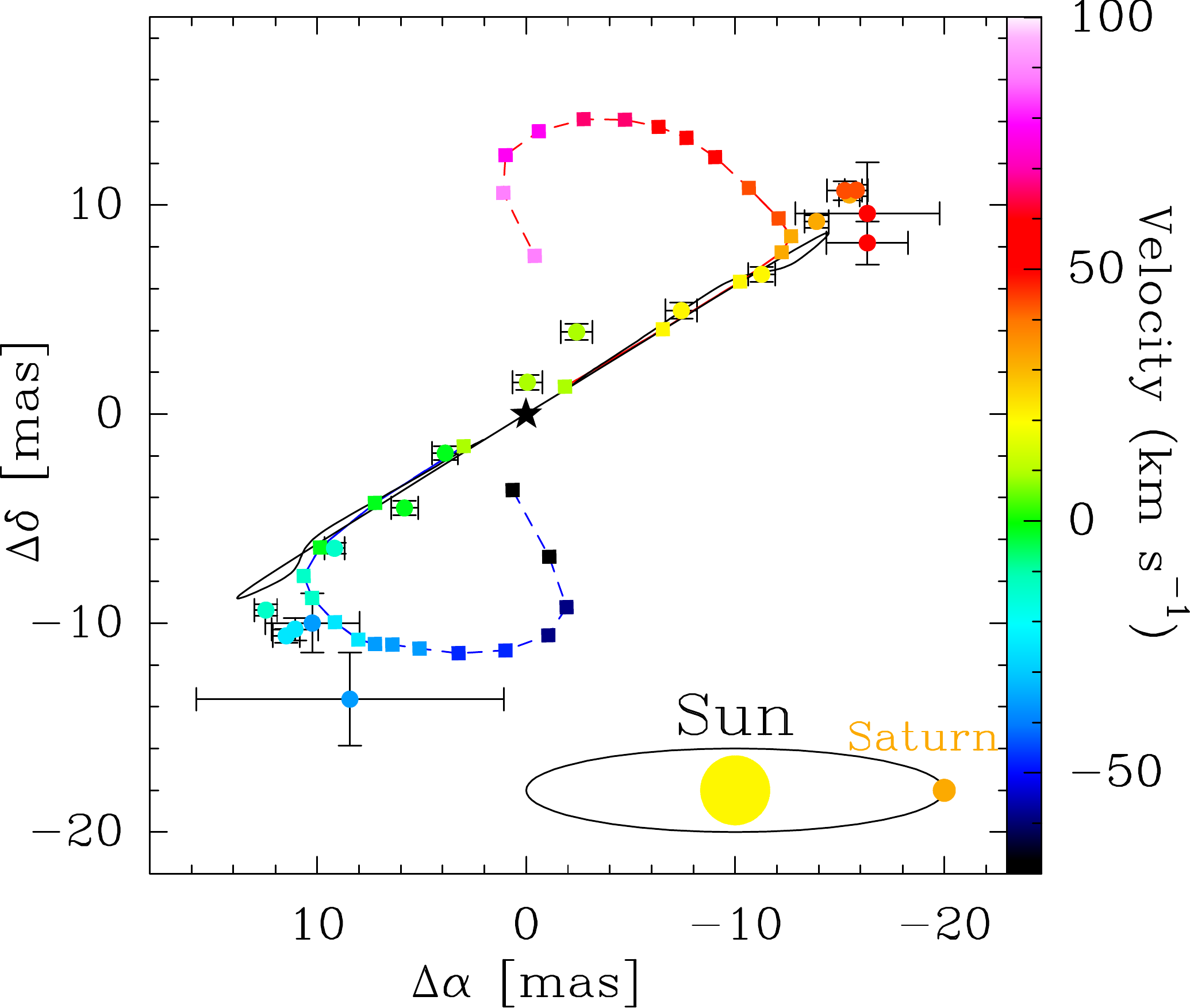}
\caption{H21$\alpha$ centroid map obtained toward MonR2-IRS2 after subtracting the central coordinates of the source. Crosses and filled circles represent the derived 2D Gaussian centroids, and the color scale shows the velocity associated with each centroid position. Error bars correspond to the $\pm$1$\,$$\sigma$ positional error from the centroid fitting. Color lines and squares report the H21$\alpha$ centroid positions and velocities predicted by the MORELI code assuming an ionized disk and disk wind for the physical structure of MonR2-IRS2. Dashed lines show the model predictions for the velocity ranges where the signal-to-noise ratio in the ALMA data is low. Black line reports the MORELI predictions for the same model but without a wind. 10$\,$mas corresponds to 9$\,$au, the Sun-Saturn distance.}
\label{centroid_map_figure}
\end{center}
\end{figure}

The bright emission of the H21$\alpha$ masers allow us to investigate the kinematics and internal structure of MonR2-IRS2 with an exquisite accuracy. For this, we have constructed the centroid map of the H21$\alpha$ emission in 5$\,$km$\,$s$^{-1}$ channels by using 2D Gaussians (Figure$\,$\ref{centroid_map_figure}, black crosses and filled circles). 
The typical errors in the centroid positions (error bars in Figure$\,$\ref{centroid_map_figure}) are 1-2$\,$mas (0.9-1.8$\,$au) for velocities between $-$20 and 40$\,$km$\,$s$^{-1}$. Note that positional errors are also a function of the errors in the bandpass calibration \citep{zhang17}. However, the phase noise in the ALMA bandpass data was $\leq$1$^\circ$, which implies a negligible effect in the positional errors of the H21$\alpha$ centroids.

Figure$\,$\ref{centroid_map_figure} shows that for velocities between $-$8 and 32$\,$km$\,$s$^{-1}$, the H21$\alpha$ centroids are distributed along an almost perfect line in the northwest-southeast direction. The centroids, however, depart from this line at high-velocities, especially for the blue-shifted gas ($\leq$$-$8$\,$km$\,$s$^{-1}$), although some of the associated positional errors are large. This suggests the presence of a second kinematic component. As shown in Section$\,$\ref{model}, this configuration is consistent with the presence of an almost edge-on ionized disk, and a disk wind. 

\section{Modelling of the H21$\alpha$ RRL}
\label{model}

\begin{figure}
\begin{center}
\includegraphics[angle=0,width=1.0\textwidth]{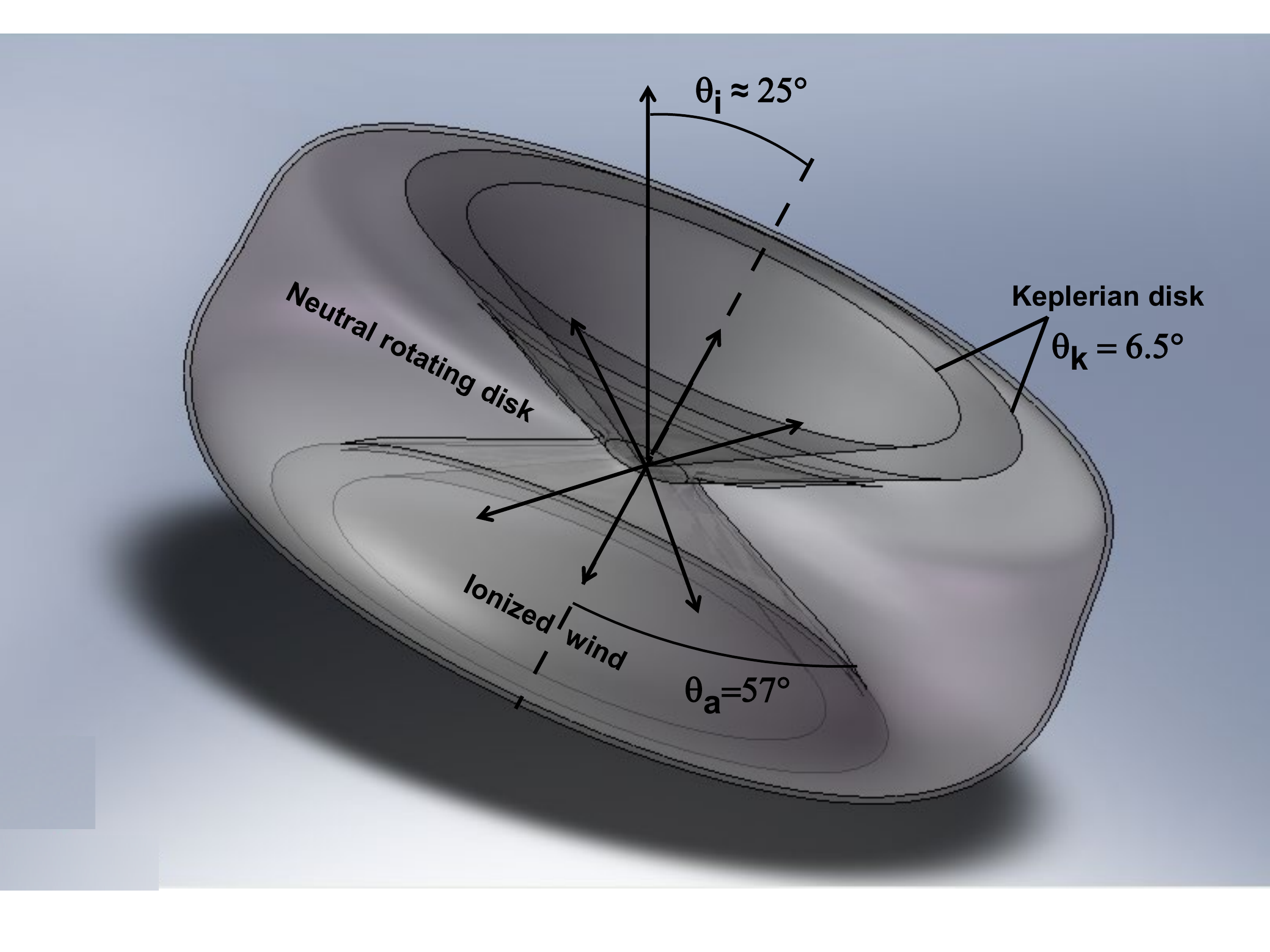}
\caption{Sketch of the double-cone geometry used for the modelling of MonR2-IRS2 \citep[see also][]{martin-pintado11,baez13}.}
\label{fig4}
\end{center}
\end{figure}

Similar configurations to the RRL centroid map obtained for MonR2-IRS2 have been observed for MWC349A \citep{martin-pintado11,baez13,zhang17} and MWC922 \citep[][]{sanchez-contreras19}. This configuration has been modelled assuming that the massive star is surrounded by a (partly) ionized Keplerian disk and an ionized wind. We have thus adopted a similar physical structure to model the H21$\alpha$ centroid map derived for MonR2-IRS2. The set of input parameters that best fit our ALMA observations are described below (see also \mbox{Table \ref{inputparameters}}).

In our model, the ionized wind lies inside a double-cone with a semi-opening angle of 57$\degr$ (the neutral disk), while the ionized disk layer is contained within an opening angle of 6.5$^\circ$ on the surface of the double-cone (Figure$\,$\ref{fig4}). The ionized wind expands radially at a constant velocity, $v_0$, while the circumstellar disk rotates following a Keplerian law with the velocity rotation component being added to the expansion velocity of the wind \citep{martin-pintado11,baez13}. The inclination angle is measured from the direction perpendicular to the disk and is $-$25$^\circ$.

We have used the non-LTE 3D radiative transfer code MORELI \citep[MOdel for REcombination LInes;][]{baez13} and the departure coefficients of hydrogen, $b_n$ and $\beta_n$, calculated by \citet{walmsley90}. The velocity separation between the H21$\alpha$ emission peaks (blue- and red-shifted; Figure$\,$\ref{spectrum}) constrains the central mass of the MonR2-IRS2 source to 15$\,$M$_\odot$. This mass corresponds to a B0-type star on the zero-age main sequence (ZAMS), which is
consistent with the luminosity of MonR2-IRS2 \citep[$\sim$0.5-1$\times$10$^4$$\,$L$_\odot$;][]{howard94}.

The radial density distribution of electrons within the ionized double-cone, $N_e$, follows a $r^{-2}$ law with an inner radius of 13$\,$au (Table$\,$\ref{inputparameters}). We choose a $r^{-2}$ law because it fits nicely the 1.3$\,$mm, 0.85$\,$mm and 0.4$\,$mm free-free continuum fluxes of MonR2-IRS2, and it allows us to simultaneously model the disk and the expanding wind in a simple way.
The predicted mass loss rate for MonR2-IRS2 is 2.8$\times$10$^{-7}$$\,$M$_\odot$$\,$yr$^{-1}$, and the mass of the ionized disk is 2.4$\times$10$^{-5}$$\,$M$_\odot$, a factor $\geq$10 lower than the neutral disk mass ($\sim$2-4$\times$10$^{-4}$$\,$M$_\odot$; Section$\,$\ref{continuum}).

The electron temperature, T$_e$, that best fits our ALMA data is 5850$\,$K. 
This value is lower than those typically measured in HII regions ($\sim$8000$\,$K). However, note that T$_e$ can only be constrained by using RRL thermal emission. The expanding velocity of the ionized wind is 10$\,$km$\,$s$^{-1}$ with a turbulent component of 5$\,$km$\,$s$^{-1}$. Expansion velocities $\geq$20$\,$km$\,$s$^{-1}$ are ruled out because the predicted H21$\alpha$ linewidths in the wings would be too large. As shown below, the absence of an expanding wind cannot explain the departure of the H21$\alpha$ centroids from the disk plane observed at high velocities (especially blue-shifted).  

Our model fits very well the H21$\alpha$ RRL profile measured toward MonR2-IRS2 (red line, Figure$\,$\ref{spectrum}). The comparison between the LTE and non-LTE cases\footnote{The LTE spectrum is obtained by fixing $b_n$ and $\beta_n$ to 1.} reveals that non-LTE effects need to be invoked to explain the H21$\alpha$ double-peaked profile. For the H30$\alpha$ and H26$\alpha$ RRLs, the model predicts peak intensities a factor of $\sim$2 higher than observed. However, we note that RRL masers are highly variable \citep[][]{thum92}.

The H21$\alpha$ centroid distribution predicted by our model also nicely matches that measured with ALMA (color lines and filled squares; Figure$\,$\ref{centroid_map_figure}). The Keplerian component reproduces fairly well the kinematics of H21$\alpha$ emission along the disk plane. The observations, however, reveal small deviations with respect to the model at velocities 2$\,$km$\,$s$^{-1}$, 12$\,$km$\,$s$^{-1}$, and 17$\,$km$\,$s$^{-1}$ (such deviations are larger than the centroids positional errors), as if the disk were warped. Disk warping may appear in systems presenting a secondary object such as a stellar companion or a massive planet \citep{nealon18,kraus20_1}. This idea would be consistent with the fact that the modelled electron density distribution of MonR2-IRS2 requires an internal hole of radius 13$\,$au. This was already noted by \citet{jimenez-serra13} since, unlike MWC349A, the free-free continuum emission of MonR2-IRS2 at $\lambda$$\leq$1.3$\,$mm is optically-thin. Alternatively, the warping of the disk could be produced by anisotropic accretion of gas \citep[][]{sakai19} or by a misalignment between the disk rotation axis and the magnetic field direction \citep{ciardi10}. 

From Figure$\,$\ref{centroid_map_figure}, we also find that the ionized gas at high-velocities (mostly blue-shifted) departs from the disk plane at distances $\sim$12$\,$mas from the disk center. This departure can only be explained by the presence of an ionized wind launching at radii $\sim$11$\,$au. Indeed, the same model but without a wind, does not reproduce such deviations (black line in Figure$\,$\ref{centroid_map_figure}). As found toward other sources \citep[MWC349A or MWC922;][]{martin-pintado11,baez13,zhang17,sanchez-contreras19}, launching radii of $\sim$10$\,$au are consistent with magnetically-regulated disk wind models \citep[][]{blandford82} rather than with X-wind theory \citep[][]{shu94}. Disk UV photoevaporation cannot be responsible either for the formation of the disk wind in MonR2-IRS2 because its gravitational radius, r$_g$ \citep[the distance at which photoevaporating disk winds are launched;][]{hollenbach94}, is $\sim$100$\,$au, i.e. ten times larger than observed in MonR2-IRS2.  

Alternatively, the departure of the high-velocity centroid positions could be due to a second warp in the outer disk of MonR2-IRS2. However, as shown by \citet{nealon18}, the disk twist produced by a companion such as a massive planet, can go up to 60$\,$$\degr$. This value is smaller than the putative disk twist observed for MonR2-IRS2 ($>$90$\,$$\degr$, Figure$\,$\ref{centroid_map_figure}).

We finally note that the model used here is different from the one proposed by \citet{jimenez-serra13}. The previous model was motivated by the asymmetry of the H26$\alpha$ RRL detected toward MonR2-IRS2 with the SMA. Such an asymmetry is not observed in the H21$\alpha$ line (Section$\,$\ref{h21a}), which suggests that the H26$\alpha$ maser may be variable. Variability is common in RRL masers since their line profiles can change in time-scales as short as 30$\,$days \citep[see MWC349A;][]{martin-pintado89b,thum92}. Future ALMA observations will reveal whether the H26$\alpha$ line asymmetry detected with the SMA was transitory or not.

\section{Conclusions}
\label{conclusion}

Sections$\,$\ref{results} and \ref{model} show that RRL masers are powerful tools to unveil the kinematics and physical structure of the innermost regions around massive stars. Indeed, RRL masers are very sensitive to the electron density distribution, geometry, electron temperature and kinematics of the ionized gas \citep[][]{strelnitski96,baez13,baez14}. Regardless their evolutionary stage, all RRL maser objects firmly detected to date \citep[MWC349A, MWC922 and MonR2-IRS2;][and this work]{martin-pintado11,baez13,sanchez-contreras19}, present the same physical structure for the ionized gas: a circumstellar disk rotating in a Keplerian fashion, and an expanding ionized wind launched at distances of tens of au that follows the same Keplerian rotation. 
This disk+wind geometry may thus be the optimised configuration for RRL masers to form. 

\begin{deluxetable}{lc}
\tablecaption{Input parameters of the best fit of the MonR2-IRS2 H21$\alpha$ data}
\tablewidth{0pt}
\tablehead{
\colhead{Parameter} & \colhead{Value} }
\startdata
Central mass, $M_*$    	     & 15$\,$$M_\odot$  \\ 
Density distribution\tablenotemark{a}, $N_\mathrm{e}(r,\theta)$ & 5.36$\times$10$^7$r$^{-2}$ $\exp{\left[\left(\theta-\theta_\mathrm{a}\right)/20\right]} $ $\ \mathrm{cm}^{-3}$\\
Inner radius, $r_\mathrm{min}$ & 13$\,$au \\
Double-cone's semi-opening, $\theta_\mathrm{a}$ & 57$\degr$ \\
Electron temperature, $T_\mathrm{e}$ & 5850$\,$$\mathrm{K}$\\
Inclination angle, $\theta_\mathrm{i}$ &
 -25$\degr$ \\
Opening angle of the ionized disk, $\theta_\mathrm{d}$ & 6.5$\degr$  \\
Outflow terminal velocity, $v_0$ & 10$\,$km$\,$s$^{-1}$\\
Outflow turbulent velocity, $v_\mathrm{tu}$ & 5$\,$km$\,$s$^{-1}$ \\
Mass loss rate, $\dot{M}$ & 2.8$\times$10$^{-7}$$\,$$M_\odot$$\,$yr$^{-1}$ \\
Mass of the ionized disk, $m_{d}$ & 2.4$\times$10$^{-5}$$\,$$M_\odot$
\enddata
\tablenotetext{a}{$r$ is in units of 15$\,$au and angles are measured relative to the polar axis; $\theta$ and $\theta_\mathrm {a}$, are in degrees.}
\label{inputparameters}
\end{deluxetable}

\acknowledgments

We thank an anonymous referee for his/her constructive comments. This paper makes use of the ALMA data ADS/JAO.ALMA 2012.1.00522.S. ALMA is a partnership of ESO (representing its member states), NSF (USA) and NINS (Japan), together with NRC (Canada), NSC and ASIAA (Taiwan), and KASI (Republic of Korea), in cooperation with the Republic of Chile. The Joint ALMA Observatory is operated by ESO, AUI/NRAO and NAOJ. I.J.-S., A.B.-R. and J.M.-P. acknowledge support from the Spanish FEDER through project ESP2017-86582-C4-1-R. V.M.R. has received funding from the EU Horizon 2020 programme under the Marie Sk\l{}odowska-Curie grant agreement No$\,$664931. We thank A. Richards for the help provided during the calibration/imaging of the data. 



\end{document}